\documentclass[5p]{elsarticle}

\usepackage{lineno}

\title{Status and Strategies of Current LUNASKA Lunar Cherenkov Observations with the Parkes Radio Telescope}

\author[adelaide,atnf]{J.D. Bray\corref{corresponding}}
\author[atnf]{R.D. Ekers}
\author[atnf]{P. Roberts}
\author[atnf]{J.E. Reynolds}
\author[nijmegen]{C.W. James}
\author[atnf]{C.J. Phillips}
\author[melbourne]{R.A. McFadden}
\author[adelaide]{R.J. Protheroe}
\author[adelaide]{M. Aartsen}
\author[santiago]{J. Alvarez-Mu\~{n}iz}

\cortext[corresponding]{Corresponding author: \texttt{justin.bray@gmail.com}}

\address[adelaide]{Dept. of Physics, School of Chemistry \& Physics, Univ. of Adelaide, SA 5005, Australia}
\address[atnf]{CSIRO Astronomy and Space Science, Epping, NSW 1710, Australia}
\address[nijmegen]{Dept. of Astrophysics, IMAPP, Radboud Univ. Nijmegen, P.O. Box 9010, 6500GL Nijmegen, The Netherlands}
\address[melbourne]{School of Physics, Univ. of Melbourne, VIC 3010, Australia}
\address[santiago]{Dept. F\'{i}sica de Part\'{i}culas \& IGFAE, Univ. Santiago de Compostela, 15782 Santiago, Spain}

\begin{document}

\begin{abstract}
LUNASKA (Lunar UHE Neutrino Astrophysics with the Square Kilometre Array) is an ongoing project conducting lunar Cherenkov observations in order to develop techniques for detecting neutrinos with the next generation of radio telescopes.  Our current observing campaign is with the 64-metre Parkes radio telescope, using a multibeam receiver with 300 MHz of bandwidth from 1.2-1.5 GHz.  Here we provide an overview of the various factors that must be considered in the signal processing for such an experiment.  We also briefly describe the flux limits which we expect to set with our current observations, including a directional limit for Centaurus~A.
\end{abstract}


\maketitle


\section{Introduction}

The lunar Cherenkov technique is a method for detecting ultra-high energy (UHE) particles which makes use of the regolith of the Moon as a target volume, giving it a larger potential aperture than any other current method.  A UHE particle shower in this material will produce a coherent radio pulse through the Askaryan effect~\cite{askaryan}, which may be observed by terrestrial radio telescopes, as proposed by Dagkesamanskii and Zheleznykh~\cite{dagkesamanskii}.  Active experiments of this type include NuMoon~\cite{buitink10}, RESUN~\cite{jaeger09}, and our own LUNASKA (Lunar UHE Neutrino Astrophysics using the Square Kilometre Array)~\cite{james10}.

The long-term goal of the LUNASKA project, as the name suggests, is to develop strategies for and eventually utilise the Square Kilometre Array: a large radio telescope currently being planned for future construction in either Australia or South Africa~\cite{ska}.  Our observations thus far have made use of the Australia Telescope Compact Array (ATCA)~\cite{james10} - and, now, the Parkes radio telescope.  Here we focus on the signal processing requirements of our current observations, the understanding of which is essential for lowering the energy threshold at which UHE particles may be detected.

\section{Observations with the Parkes Radio Telescope}

The Parkes radio telescope is a single-dish telescope with a diameter of 64 metres, located in New South Wales, Australia.  We are conducting 200 hours of lunar Cherenkov observations with it during 2010, using its 20cm multibeam receiver.  This receiver has 300 MHz of bandwidth from 1.2-1.5 GHz, and allows us to place multiple pointings or `beams' on the sky simultaneously, with dual linear polarisations.  Each beam has a full width at half maximum (FWHM) size of around 14 arcminutes.

\begin{figure}
\centering
\includegraphics[width=0.7\linewidth]{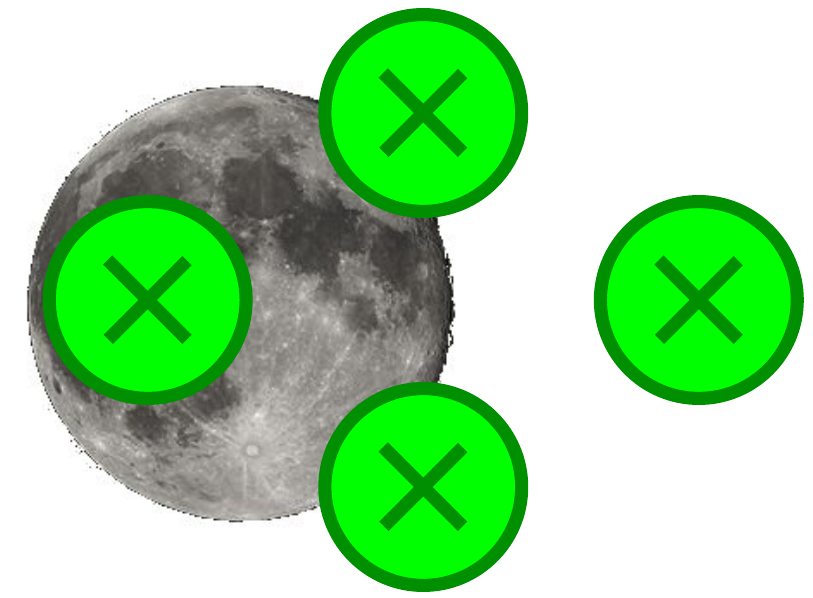}
\caption{Pointing strategy with Parkes radio telescope.  Circles represent the positions of beams with respect to the Moon, and crosses within them indicate the alignment of the linear polarisations.  The off-moon beam is for anticoincidence filtering - see section \ref{sec:anticoincidence}.}
\label{fig:pointing}
\end{figure}

A radio pulse from a lunar Cherenkov event is expected to come from the limb of the Moon, with radial polarisation, which guides our decision on where to point the beams.  In addition, the system noise in our observations is dominated by the thermal noise of the Moon, so we must compromise between pointing more on the Moon (less sensitive), or slightly off the Moon (more sensitive; but observing a smaller area).  We have explored several pointing strategies, but have spent most observing time in the configuration shown in figure \ref{fig:pointing}.  This configuration allows us to keep two beams slightly off the Moon, with their linear polarisations aligned such that the power of a single event may be detected mostly on a single polarisation channel.

The sensitivity of the lunar Cherenkov technique to UHE particles is highly directionally dependent~\cite{james09}, which makes it possible to target observations at a particular potential source, to maximise sensitivity to it.  In our observations, we have targeted the radio galaxy Centaurus~A, which is associated with an excess of UHE cosmic ray arrival directions~\cite{abreu10}.  We note that the total power of the neutrino flux from Centaurus~A is limited by gamma-ray observations~\cite{kachelriess09}; however, the contribution of the flux in our energy range to the total power is small, so it is less strongly constrained.

\section{Signal processing}

We use the standard front-end signal path for the 20 cm multibeam receiver at Parkes, with minor modifications (see section \ref{sec:attenuation}).  The receiver amplifies the signal, and mixes it with a sine-wave local oscillator at 1555 MHz to downconvert it from the 1.2-1.5 GHz radio band to 55-355 MHz.

At this point, the signal enters our purpose-built digital backend, the `Bedlam' Neutrino Transient Processor.  It is converted from analog to digital with 8 bits of precision, at a rate of 1.024 Gsamples/s.  Then, it is passed through a finite impulse response (FIR) filter, with 64 coefficients or `taps' (see section \ref{sec:dedispersion}).  The samples, both before and after filtration, are stored in a running buffer with a length of up to 8k samples.  A threshold test is applied both to these and to interpolated values (see section \ref{sec:interpolation}) and, if the appropriate anticoincidence criteria are met (see section \ref{sec:anticoincidence}), the contents of all buffers are written to disk.  While the buffers are being stored, the system is not processing incoming data: the fraction of this `dead time' was generally held to around 5\%.

Further processing (see sections \ref{sec:phase} and \ref{sec:bandshaping}) is applied to the stored buffers in post-analysis.  The real-time processing is not required to be exhaustive: it only needs to ensure that any potentially significant lunar Cherenkov event is stored, so that it can be identified later.

\subsection{Attenuation}
\label{sec:attenuation}

We tested the response of the telescope by transmitting a sine-wave from a small antenna on the telescope surface.  The received sine-wave was observed to be clipped, with its peaks truncated.  This is similar to the receiver saturation problem encountered in a similar experiment using another telescope, the Very Large Array, by Jaeger et al~\cite{jaeger09}.  In our case, the problem was traced to a particular amplifier which was being driven outside of its design voltage range, and failing to maintain linear amplification of the signal.  This was solved by attenuating the signal earlier in the signal chain, bringing the amplifier back into its design range, and using our own amplifiers to restore the signal afterwards.  Subsequent repetition of the earlier test confirmed that this had the desired effect.


\subsection{Dedispersion}
\label{sec:dedispersion}

\begin{figure}[t]
\centering
\includegraphics[width=\linewidth]{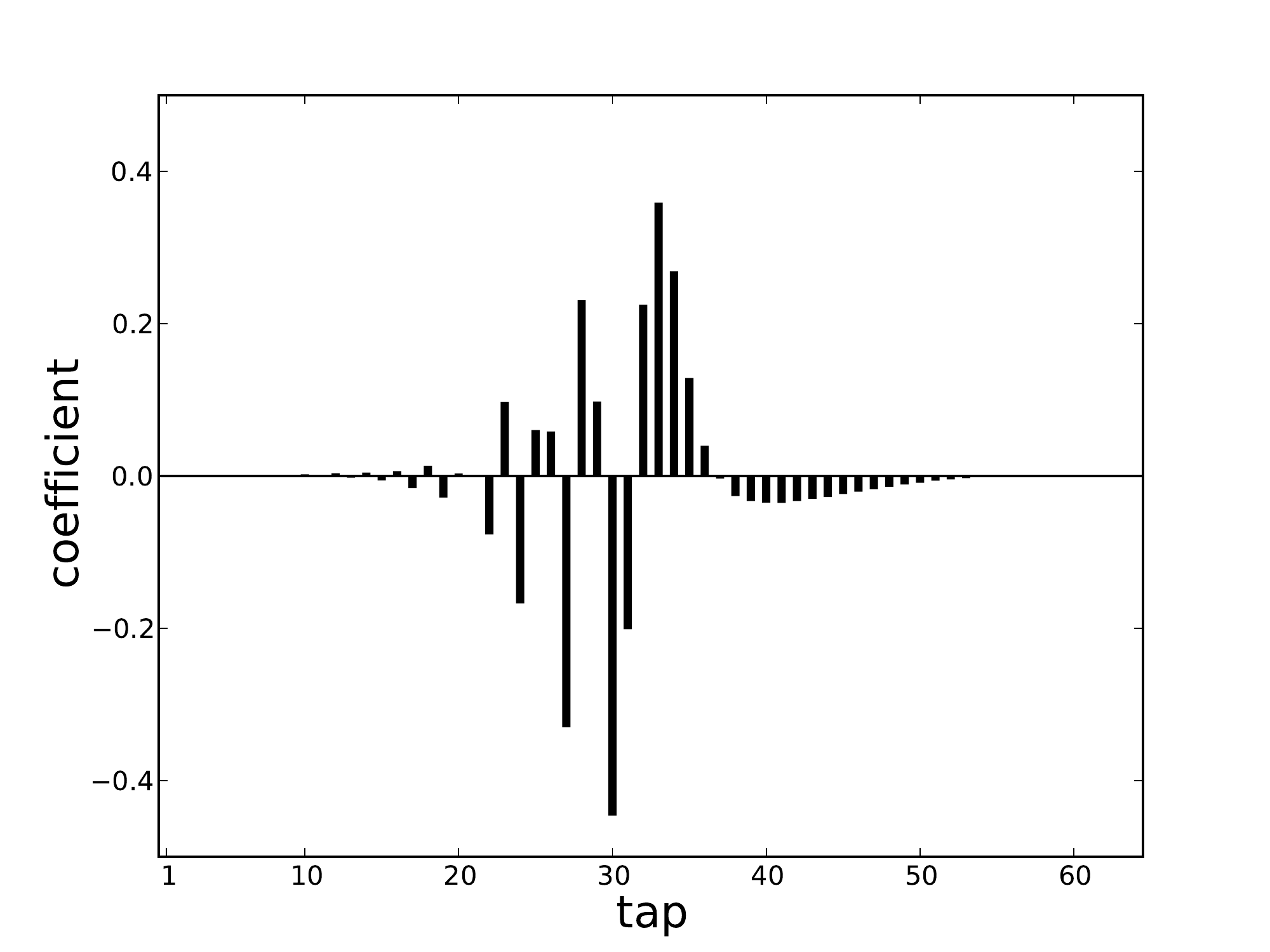}
\caption{Impulse response of de-dispersion filter compensating for 7 ns of dispersion, which is a typical maximum value during one of our observing sessions.  This filter is equivalent to correlating with an expected pulse shape the same as this impulse response.  Note the separation of high-frequency components (left) and low-frequency components (right).}
\label{fig:impulse_response}
\end{figure}

\begin{figure}[t]
\centering
\includegraphics[width=\linewidth]{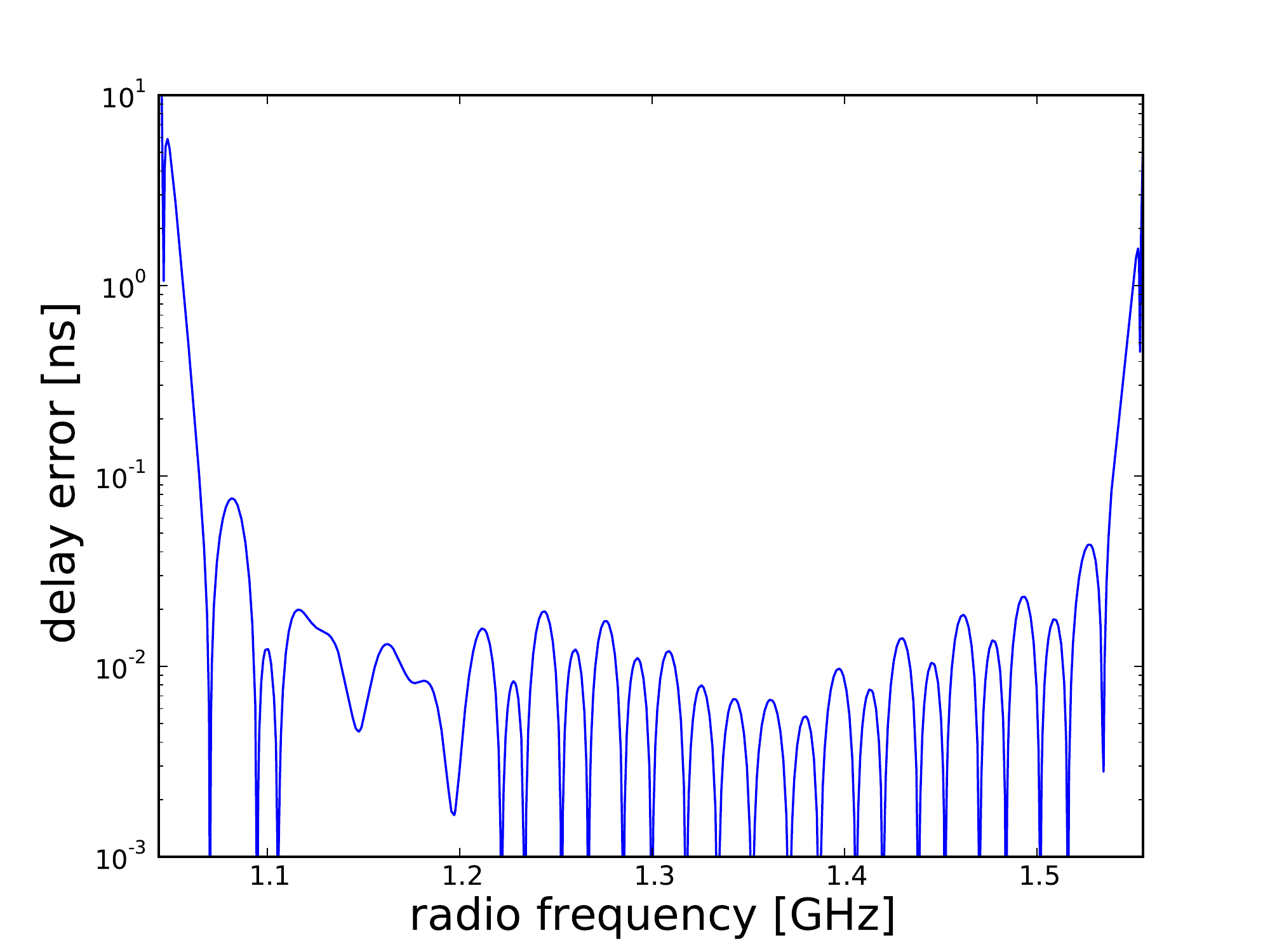}
\caption{Performance of filter shown in figure \ref{fig:impulse_response}, measured as deviation from the ideal delay at each radio frequency.  The band containing the signal is 1.2-1.5 GHz.}
\label{fig:diff_delay}
\end{figure}

It has been recognised since the first lunar Cherenkov observations by Hankins et al~\cite{hankins96} that a broadband coherent pulse from the Moon will be dispersed as it passes through the ionosphere, causing the low-frequency components to be delayed with respect to the high-frequency components, and thus reducing the peak amplitude of the pulse.  In our previous experiments with the ATCA, we used an analog de-dispersion filter to compensate for this effect.  In our current observations, we use a digital finite impulse response (FIR) filter implemented in our digital backend.  This has the advantage that the filter can be easily reconfigured to apply a different amount of dedispersion, so we can adjust it during each observing session to match changes in ionospheric conditions and the slant depth of our sight-line to the Moon.  A typical filter configuration is shown in figure \ref{fig:impulse_response}, and its performance characterised in figure \ref{fig:diff_delay}.

The parameter of the ionosphere that determines the magnitude of the dispersion is its total electron content (TEC), for which we obtain hourly figures from the Australian Bureau of Meteorology~\cite{ionosonde}, who produce ionospheric models based on ionosonde radar measurements.

\subsection{Interpolation}
\label{sec:interpolation}

If the sampling rate of the digital signal were infinite, then any pulse would always be sampled at its peak, allowing a threshold test to detect its full amplitude.  However, the finite sampling rate of any real system means that there will not generally be a sample taken exactly at the peak of the pulse, and so the full amplitude of the pulse will not be detected.  The loss of sensitivity due to this effect is described in \cite{james10}.

Since the input signal is band-limited, and sampled at over the Nyquist frequency, the Shannon-Nyquist sampling theorem tells us that our samples contain full information about the original analog signal, including any missed peaks, and the Whittaker-Shannon interpolation formula allows us to obtain missing intermediate values by convolving the signal with a \texttt{sinc} function.  This allows the full amplitude of a pulse to be recovered, and is computationally cheaper than directly increasing the sampling rate.  However, it is still somewhat computationally intensive, and becomes more so as more intermediate values are required, so our real-time hardware reconstructs only one intermediate value between each pair of real samples (thus effectively doubling the sampling rate) - which is sufficient to ensure that any significant real pulse will be stored - and more exhaustive interpolation is performed in post-processing.

\subsection{Anticoincidence}
\label{sec:anticoincidence}

A major concern for observations of this type is that there exists radio-frequency interference (RFI) that resembles nanosecond-scale pulses of the sort that would be expected from a lunar Cherenkov event.  Such pulses can be detected from the surroundings of the telescope via the far sidelobes of its beam.  However, since we are using a multibeam receiver, these pulses are distinguished by appearing on all beams simultaneously.  By contrast, a real lunar Cherenkov event should appear only on the beam pointing at the part of the Moon where it occurred.

We therefore apply an anticoincidence filter, excluding potential events which appear in more than one beam.  A simple such filter is applied in real time, excluding the majority of pulse-like RFI, and a more sophisticated version is applied in post-analysis.  The off-moon beam, having a lower system temperature, is more sensitive, and particularly effective at identifying RFI pulses.

\subsection{Phase search}
\label{sec:phase}

Recent simulations of Cherenkov pulses in the time-domain show that their shape is bipolar, with separate positive and negative peaks~\cite{romero-wolf10}.  In the frequency domain, this corresponds to the pulse being entirely imaginary, with a phase of $\pm \pi/2$.  However, the process by which a radio telescope converts a signal from radio frequency (RF) to a lower intermediate frequency (IF) will change this phase randomly, depending on the phase offset between the arriving pulse and the local oscillator (LO) signal.  We implement a search through the possible shapes of the original pulse, for different possible LO phases.

\subsection{Band shaping}
\label{sec:bandshaping}

Optimum detection of a known pulse against a background of noise requires `pre-whitening' the noise spectrum, then correlating with the expected pulse shape~\cite{prewhitening}.  In the frequency domain, this is equivalent to weighting different frequencies according to the expected signal-to-noise ratio at each of them.

The thermal noise from the Moon follows a Rayleigh-Jeans spectrum, with spectral power proportional to $\nu^2$, which is the same as the spectrum of a fully-coherent Cherenkov pulse.  Since this source dominates the total noise present in the signal, the signal-to-noise ratio is approximately constant across all frequencies, and the entire band should be weighted equally.  However, it may be desirable to weight lower frequencies more heavily to optimise for the detection of partially-incoherent Cherenkov pulses, for which the spectrum cuts off at higher frequencies.

\section{Sensitivity}

\begin{figure}
\centering
\includegraphics[width=\linewidth]{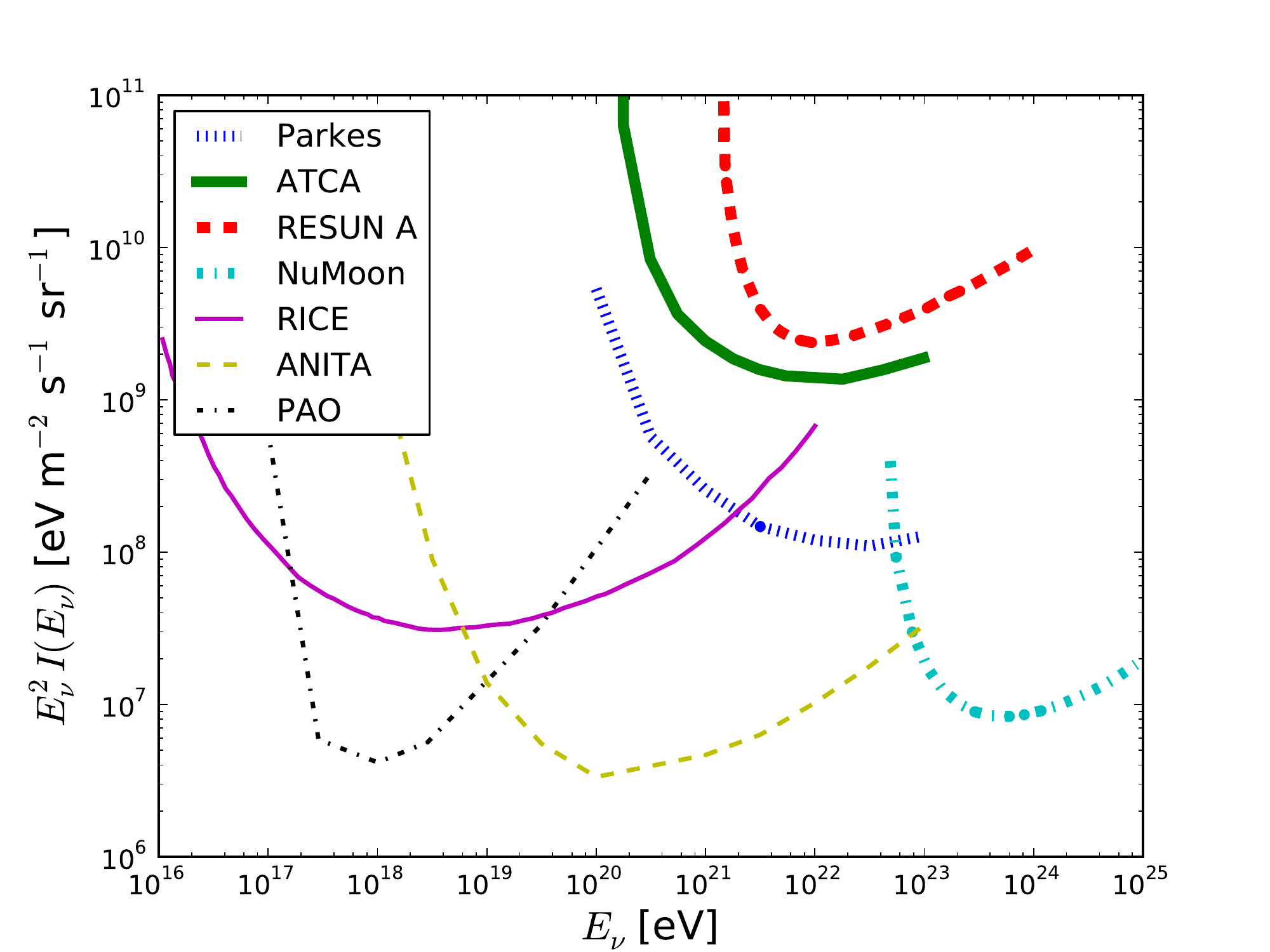}
\caption{Differential limits (90\% confidence) on the isotropic all-flavour neutrino flux set by lunar Cherenkov observations (thick lines) and other experiments (thin lines).  The curve for Parkes represents the limit we expect to set with our current observations, assuming no detections in 200 hours.  Other lunar Cherenkov observations include RESUN A~\cite{jaeger09}, NuMoon~\cite{buitink10}, and our own observations with the ATCA~\cite{james10}.  Other experiments include RICE~\cite{kravchenko06}, ANITA~\cite{gorham10}, and the Pierre Auger Observatory (PAO)~\cite{tiffenberg09} (for upgoing neutrinos only).  A 1:1:1 flavour ratio is assumed.}
\label{fig:iso_nu}
\end{figure}

\begin{figure}
\centering
\includegraphics[width=\linewidth]{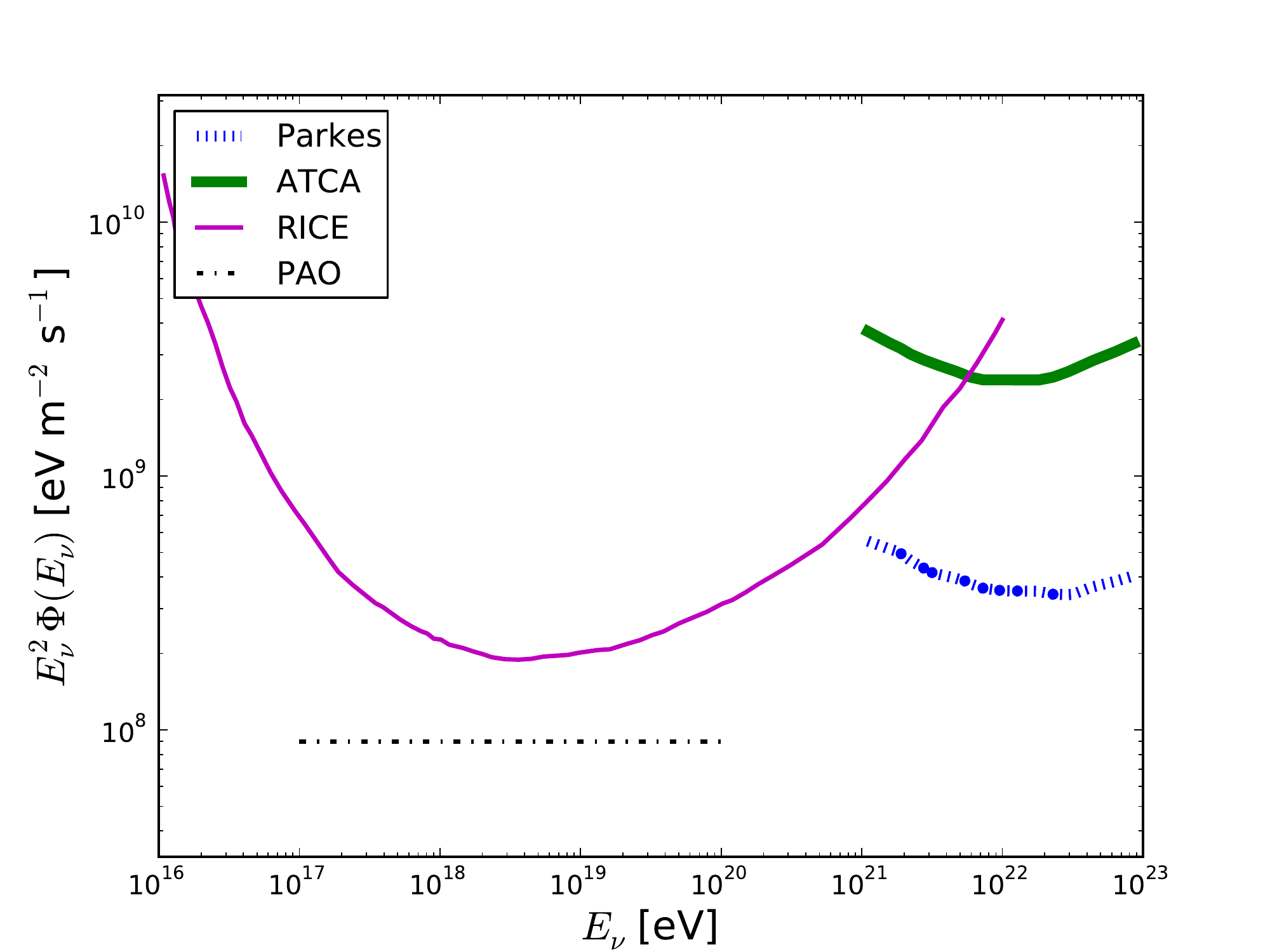}
\caption{Differential limits (90\% confidence) on an all-flavour neutrino flux from Centaurus A.  As in figure \ref{fig:iso_nu}, the curve for Parkes is for the projected limit from our current observations.  Also shown are the limit from our previous observations with the ATCA~\cite{james10b}, a limit derived~\cite{james10b} from the exposure of RICE~\cite{kravchenko06}, and a preliminary integrated (not differential) limit set by the PAO~\cite{tiffenberg09}.  Other experiments in figure \ref{fig:iso_nu} have minimal exposure to Centaurus A.}
\label{fig:cena_nu}
\end{figure}

Using the Monte Carlo simulation described in \cite{james07}, we calculate the sensitivity of our experiment, and the limits we expect to set to a flux of UHE neutrinos (assuming no detections).  For an isotropic flux, our projected limit is shown in figure \ref{fig:iso_nu}, along with limits from other experiments in the same energy range.  For a directional flux from Centaurus A, our projected limit is shown in figure \ref{fig:cena_nu}.  Note that Centaurus A has extended structure on a scale comparable to the size of the region on the sky from which we are sensitive to UHE particles, so the limit to a flux from its outer structure (i.e., its giant lobes) will be slightly different.

The version of the simulation software used here does not incorporate any variation of the gradient of the lunar surface over the scale of a particle shower.  The effect of surface roughness on this scale will generally be to reduce the aperture of the experiment at low energies, and to increase it at high energies; see appendix B of~\cite{james10} for more discussion of this effect.

\section{Conclusion}

We are currently conducting our most sensitive lunar Cherenkov experiment to date.  We are deliberately targeting Centaurus A, and expect that these observations will set the lowest limit on its neutrino flux in the energy range to which we are sensitive.  We are refining the signal-processing techniques that will be used in future experiments with more sensitive telescopes, eventually including the SKA.


\section{Acknowledgements}

This research was supported as a Discovery Project (DP0881006) by the Australian Research Council.  The Parkes Observatory is part of the Australia Telescope which is funded by the Commonwealth of Australia for operation as a National Facility managed by CSIRO.  The authors would like to thank Sergey Ostapchenko for discussions on the relation between the gamma-ray and neutrino fluxes from Centaurus~A.

\end{document}